\documentclass[useAMS, usenatbib]{mn2e}
\usepackage{graphicx}
\usepackage{amsmath}
\usepackage{amssymb}
\usepackage{color}
\usepackage{url}

\newcommand  \acc     {\ifmmode {\rm km\,s}^{-2} \else km\,s$^{-2}$\fi}

\newcommand  \ergs     {\ifmmode {\rm ergs\,s}^{-1} \else ergs s$^{-1}$\fi}
\newcommand  \ergcms   {\ifmmode {\rm erg~cm}^{-2}\,{\rm s}^{-1}
                        \else erg~cm$^{-2}$\,s$^{-1}$\fi}
\newcommand  \ergcmsA  {\ifmmode{\rm erg\,cm}^{-2}\,{\rm s}^{-1}\,{\rm\AA}^{-1}
                        \else ergs\,cm$^{-2}$\,s$^{-1}$\,\AA$^{-1}$\fi}
\newcommand  \ergcmsHz {\ifmmode{\rm ergs\,cm}^{-2}\,{\rm s}^{-1}\,{\rm Hz}^{-1}
                        \else ergs\,cm$^{-2}$\,s$^{-1}$\,Hz$^{-1}$\fi}
\newcommand  \phcms    {\ifmmode {\rm ph\,cm}^{-2}\,{\rm s}^{-1}
                        \else ph\,cm$^{-2}$\,s$^{-1}$\fi}
\newcommand  \phcmsA   {\ifmmode {\rm ph\,cm}^{-2}\,{\rm s}^{-1}\,{\rm\AA}^{-1}
                        \else ph\,cm$^{-2}$\,s$^{-1}$\,\AA$^{-1}$\fi}
\newcommand\aj{{AJ}}% 
\newcommand\araa{{ARA\&A}}% 
\newcommand\apj{{ApJ}}% 
\newcommand\mnras{{MNRAS}}% 
\newcommand\prl{{Phys.~Rev.~Lett.}}% 
\newcommand\nat{{Nature}}% 
\newcommand\physrep{{Phys.~Rep.}}% 
\title[Collisional Supernovae at Galactic Centers]
{A new rare type of supernovae: \\hypervelocity stellar collisions at galactic centers}

\author[S. Balberg, R. Sari \& A. Loeb]
{Shmuel Balberg$^{1}$, Re'em Sari$^{1}$ and Abraham Loeb$^{2}$\\
$^{1}$Racah Institute of Physics, The Hebrew University University, Jerusalem 91904, ISRAEL\\
$^{2}$Institute for Theory and Computation, 
Harvard University, Cambridge, MA 03210, USA}
\date{\today}

\begin{document}

\maketitle

\label{firstpage}

\begin{abstract}
When a binary star system is tidally disrupted by a supermassive black
hole at a galactic nucleus, one star is ejected at a high speed while
the other remains in a tightly bound orbit around the black hole. The
cluster of tightly bound stars builds over time, eventually creating a 
steady state in which the rate of collisions between these stars is similar 
to the rate of capturing new stars.  A large fraction of the 
collisions occur near the periapsis of the orbits around the black hole, 
where the kinetic energies are sufficient to generate an explosive disruption 
of the two stars involved.  The
typical flare brightens for several days, with a peak
luminosity that is comparable to the lower-luminosity end of known
supernovae. The explosion lightcurve is followed by a longer flare due
to accretion of ejected matter onto the black hole. Dedicated
searches in the near universe could observe several such
``collisional-supernovae'' per year.
\end{abstract}


\section{Introduction}

The typical energy output of supernova explosions, $10^{51}\;$ergs,
results in the disruption of the progenitor star. There are three
known mechanisms for supernovae: thermonuclear explosion of a white
dwarf \citep{TNSN}, core-collapse in a massive star \citep{MassiveSN}, and
pair-instability explosion of a very massive star \citep{BRS}. In this {\it Letter} we call attention to a fourth
mechanism - complete disruption of two stars following a collision at
a very high relative speed.

In order to generate energies of order $10^{51}\;$ergs, two sun-like stars must collide head-on with a relative velocity around ten thousand kilometers per second. This is much larger than typical stellar velocities in a galaxy like the Milky-Way, where collisions generally result in mergers \citep{FreitagBenz05}. However, high velocities become the standard near the supermassive black hole (SMBH) which lies at the center of the galaxy \citep{SgrA1,SgrA2}, so disruptive stellar collisions are common in galactic nuclei.
These can include very high (``hyper'') velocity collisions with an energy content at the supernovae scale. 

Observations suggest that there exists a constant supply of stars to a galactic center. This indication comes from the discovery of hypervelocity stars (HVSs) which are found to be leaving the Milky-Way with typical speeds of a few hundred to a thousand kilometers per second \citep*{HVSs1,HVSs2}. The most likely production mechanism is tidal disruption of a tight binary star system \citep{Hills88}, which approaches too close to the SMBH. The three body interaction disrupts the binary: one star is ejected with a very high velocity, while the other remains captured in a tight orbit around the SMBH. Gravitational scatterings %with other stars 
continuously supply new binaries into radial orbits towards the galactic center, and the tight-binary disruption rate is estimated theoretically to be of order one per $10^4-10^5$ years \citep*{PHT07}. \citet{Bromleyal12} have recently showed that this estimate is roughly consistent with the observed population of HVSs, as well as with the observed cluster of about twenty massive S-stars \citep*{Genzelal10} which orbit the SMBH in the galactic center.

In this {\it Letter} we quantify two implications of the above
processes. First, we show that the rate of collisions between tightly
captured stars is large enough to settle on a steady state with the
supply rate. Second, the disruptive collision gives rise to a
bolometric light curve which is on par with the lower end of observed
conventional supernovae.  This implies that the rate of potentially
observable ``collisional-supernovae'' should be significant.
 
\section{The rate of hypervelocity collisions at galactic nuclei}

A binary is disrupted if it approaches the SMBH within the tidal disruption radius, $r_T$. The captured star becomes bound to the SMBH with an orbit which has $r_T$ as its periapsis, and a larger semi-major axis, $r_1$. Both $r_T$ and $r_1$, as well as the ejection velocity of the HVS, $v_H$, all depend on the mass of SMBH, $M_\bullet$, the masses of the binary stars, $m_1$ and $m_2$, and the initial binary separation distance, $a_B$. Specifically, $r_1$ and $r_T$ can be expressed as
\begin{equation}\label{eq:r_1_and_r_T}
r_1\approx \frac{1}{2}\left(\frac{M_\bullet}{m_\star}\right)^{2/3}a_B\;\;,\;\; 
r_T\approx \left(\frac{M_\bullet}{m_\star}\right)^{1/3}a_B
\end{equation}
where we assumed for simplicity $m_1=m_2=m_\star$. Numerical simulations \citep{Hills88,Bromleyal06,SKR10} suggest that both relations are accurate up to a factor of order unity, which we neglect in the following. The observed velocity of an HVS, $v_H$, must be a few tens of percents smaller than the original velocity $v_{ej}$, due to deceleration in the gravitational field of the galaxy. Correspondingly, 
for sun-like stars, $m_\star=M_\odot$, and observed Milky Way values of $M_\bullet\approx 4\times 10^6\;M_\odot$ and $v_H\approx 10^3{\rm km}\;{\rm s}^{-1}$ the original binary separation distance must be of order $a_B=0.1$AU. Such a separation distance leads to $r_1\approx 2\times 10^{16}$cm and $r_T\approx 2.4\times 10^{14}$cm, implying very eccentric orbits. Stellar collisions near this $r_T$ will indeed be at ``hyper'' velocities, exceeding $10^9\;{\rm cm}\;{\rm s}^{-1}$.

The total rate of collisions between captured stars, $\dot{N}_{Coll}$, involves collisions at various distances $r$ from the SMBH. Assuming spherical symmetry and defining an impact parameter, $b$, the partial collision rate at each $r$ is
\begin{equation}\label{eq:N_{Coll}_r}
\frac{d\dot{N}_{Coll}(r)}{dr}=4\pi r^2 n^2(r) v_{rel}(r) \pi b^2 \;,
\end{equation}
where $n(r)$ is the number density of stars at $r$, and $v_{rel}(r)\approx (GM_\bullet/r)^{1/2}$ is the typical relative velocity of the stars. 

Now suppose that there are $N_C$ tightly captured stars in the galactic center, all with mass $m_\star$ and common values of $r_1$ and $r_T$. We assume that gravitational scatterings (dynamical relaxation) do not significantly change the orbit of each star prior to a collision - we validate this assumption below. If so, the effective number density profile $n(r)$ for $r_T\leq r \leq 2r_1$ is proportional to the time spent by a single star around $r$, divided by the volume enclosed by that radius, i.e., $n(r)\sim (r^3/GM_\bullet)^{1/2}/r^3 \sim r^{-3/2}$. 
Therefore,
\begin{equation}\label{eq:n(r)}
n(r)\sim \frac{N_C}{r^3_1}\left(\frac{r}{r_1}\right)^{-3/2}\;.
\end{equation}

Including numerical factors in equations  (\ref{eq:N_{Coll}_r}) \& (\ref{eq:n(r)}) and integrating from $r$ to $r_1$ yields a total collision rate of
\begin{equation}\label{eq:N_{CollFin}}
\begin{split}
\dot{N}_{Coll}\approx & 0.14 \frac{N^2_C}{r^3_1}\left(\frac{GM}{r}\right)^{1/2} b^2\approx \\
&  1.2\times 10^{-4}\left(\frac{N_C}{5\times 10^3}\right)^2\left(\frac{M_\bullet}{4\times10^6 M_\odot}\right)^{1/2}\times\\
& \left(\frac{M_\bullet/m_\star}{4\times 10^6}\right)^{-13/6}\left(\frac{a_B}{0.1{\rm AU}}\right)^{-7/2}\left(\frac{b}{R_\odot}\right)^2{\rm yr}^{-1}\;.
\end{split}
\end{equation}
%{\bf RS I changed$ r_T$ to $r$ in the first part and changed the text below accordingly. Also, the second part has $M_{\bullet}$ twice}.
The first part of equation (\ref{eq:N_{CollFin}}) implies that collisions near pericenter dominate the total collision rate between tightly captured stars. %This is due to the fact that the $r^{-3}$ scaling for the square of the number density cancels out with the volume enclosed by $r$, which scales as $r^3$. The $r-$dependence of the collision rate is therefore determined by the $r^{-1/2}$ factor of the relative velocity, and 
For the large eccentricity of the orbits, $r_T/r_1\sim 10^{-2}$, the majority of the collisions between the tightly captured stars occur close to $r_T$, at the highest relative velocities.

The second part of equation (\ref{eq:N_{CollFin}}), where we substituted $r \sim r_T$, presents a quantitative estimate of the hypervelocity collision rate. The critical feature is that about $5000$ tightly captured stars are required to establish a steady state with a capture rate of $10^{-4}\;{\rm yr}^{-1}$ (one collision per two stars captured). About fifteen hundred stars will suffice to create an equilibrium with a capture rate of $10^{-5}\;{\rm yr}^{-1}$. Over the lifetime of the galaxy there would have been more than $10^4$ captured stars, and so the system must settle to a steady state population, with the collision rate stabilizing at one half of the capture rate.

While this analysis ignores several process which derogate the hypervelocity collision rate, the order of magnitude should be reasonable. The most obvious issue is grazing collisions, with $R_\star<b<2R_\star$.  Such collisions may affect the colliding stars significantly, but not necessarily disrupt them. Equation (\ref{eq:N_{CollFin}}) is therefore an overestimate of the disruptive collision rate, but by a factor of four at the most. Collisions with impact factors greater than $2R_\star$ should not affect the integrity of the stars, since relative velocities are greater than the escape speeds of the stars, and gravitational effects in close encounters with $b>2R_\star$ are small.

Two-body gravitational scatterings (encounters with $b>2R_\star$) will tend to circularize the orbits which are initially highly eccentric, thus increasing the typical pericenter and reducing the collision rate at the largest velocities.
To significantly change its periaps $r_T$, the star has to gain angular momentum per unit mass of order $v_{rel}(r_T)r_T$, thus requiring a velocity change while the star is around $r_1$ of order $\delta v \sim v_{rel}(r_T)r_T/r_1$. The cross section for such an encounter is $r_1^3R_T^{-1}(m_*/M_\bullet)^2$. This process competes with the physical collisions, which have a cross section of order $R_\star^2$, but occur by at a rate higher by $(r_/r_T)^{1/2}$, because collisions close to $r_T$ dominate. Therefore, the ratio between the physical collision rate to periaps change rate is given by:
\begin{equation}
{R_*^2 (r_1/R_T)^{1/2} \over r_1^3R_T^{-1}(m_*/M_\bullet)^2}%={R_*^2 \mu^{+1/6} \over a^2 \mu^{-4/3}  \mu^2}
=2^{5/2}\left( R_* \over a \right) ^2 \left(M_\bullet \over m_* \right)^{1/2} \sim 25 
\end{equation}
We therefore find that collisions typically occur before any significant change in the periaps distance, and so the density profile should approximately maintain the $r^{-3/2}$ profile, ensuring that most collisions are of hypervelocity as
we argued. The ratio of these rates is, however, not too far above unity, considering that small angle scatterings actually reduce the periapsis change rate by a logarithmic factor of a few (i.e., $log(0.4N_C)$), and that there may be farther enhancement of the relaxation processes, e.g. by resonant relaxations or by the influence of the more massive stars.
On the other hand, binaries with separation smaller than $0.1$AU are more immune to this problem. We conclude that our estimate for the hypervelocity collision rate should be reasonable, although exact analysis through N-body simulations is required to examine this point to greater accuracy.

Another issue is that our analysis underestimates the number of lower-energy collisions the tightly bound stars can experience. These stars may collide far out from their periapsis also with other stars which are present there, but do not penetrate all the way to $r_T$. Most of these other stars will have been captured in the disruption of wider binaries and therefore have greater periapses.
For all captured stars, the periapsis distance and the semi major axis of the orbits are linearly dependent on the original separation distance, so all stars have eccentric orbits with $r_T/r_1\approx (m_\star/M_\bullet)^{1/3}$, about $10^{-2}$ for the Milky Way SMBH. Hence, the analysis presented above for the tightly bound stars applies to all captured stars of a given $a_B$. 
Consider a population of binaries with a wider separation distance, $a_{BW}$: they will lead to a cluster of captured stars near the SMBH, and if there are $N_{CW}$ such captured stars, they collide among themselves at their own periapsis at a rate $\dot{N}_{wide}$ proportional to $N_{CW}^2/a_{WB}^{7/2}$. If the periapsis of these stars is inside the orbit of the tightly captured stars, the latter will collide with the former at a rate $\dot{N}_{mix}$ proportional to $N_{CW} N_C/(a_{WB}^2 a_B^{3/2})$. For the rate $\dot{N}_{mix}$ to be comparable to the rate at which tightly bound stars collide among themselves (Eq.~(\ref{eq:N_{CollFin}})), the number of captured stars from the wider binaries must be at least $(a_{BW}/a_B)^2$ times larger than $N_C$. If so, then $\dot{N}_{wide}$ must be even higher than $\dot{N}_{mix}$, by a factor of $(a_{BW}/a_B)^{1/2}$. The implication is that stars captured from wider binaries reach a steady state among themselves before they can affect the collision rate of the tightly bound stars. Note that for the tightly bound stars with $r_T\sim 10^{16}$cm, any collision along their orbits is at sufficiently high velocities to be disruptive. Correspondingly, for wide stars which can collide with the tightly bound population, $\dot{N}_{wide}$ is essentially the rate of disruptive collisions of the wider population.

The distribution of binary separation distances is observationally inferred to be logarithmic, i.e., $P(a_B)da_B\sim da_B/a_B$ \citep{binaries1,binaries2}. The capture rate should therefore be independent of binary separation distance (and even a slight enhancement for larger separation distances due to a logarithmic factor of an ``empty loss cone'' distribution \citep{FrankRees76,LightmanShapiro77}). The immediate conclusion is that stars from wider binaries cannot accumulate to a number that will affect the collision rate of the tightly bound stars, without first reaching a steady state between their capture rate and being destroyed by collisions among themselves. In reality, the wider population must therefore be too small to significantly affect the hypervelocity collision rate, either because they have not accumulated to a significant number of their own, or that collisions among themselves have limited their population to an equilibrium with their own capture rate. We conclude that stars captured from wider binaries do not pose a significant threat to the hypervelocity collision rate.

It is noteworthy that single stars also scatter gravitationally towards the galactic center. Those which approach the black hole as close as their tidal radius are disrupted, but others complete multiple orbits as they diffuse in momentum space until finally having a periapsis which is too large to be of interest. These stars might collide with the tightly bound captured stars, contributing to hypervelocity collisions if the collision occurs near $r_T$, but harming their rate if the collision occurs farther out. Single stars which cross a radius $r$ from the SMBH at a rate of $\dot{N}_s(r)$, collide with tightly captured stars with a semi-major axis of $r_1=r$ at a rate of $\sim \dot{N}_s(r)N_C(r)\times (R_\star/r)^2$. For each original large distance $x_0$ from the SMBH, the rate at which stars that originate around $x_0$ cross a spherical surface at a distance $r$ from the SMBH can be estimated with a straightforward approximation. Assuming an isotropic distribution of velocities at $x_0$, this rate corresponds to the ``full loss cone'' limit   
\begin{equation}\label{eq:N_dot_single}
\dot{N}_s= 4\pi x^2_0 n(x_0) \sigma(x_0) \frac{1}{2}\frac{r_1}{x_0}\;,
\end{equation} 
where $n(x_0)$ and $\sigma(x_0)$ are the number density and velocity dispersion of stars at $x_0$. For typical Milky Way values, the dominant contribution to small angle scattering of stars which approach the SMBH occurs at $x_0\approx 1$pc, with $n(x_0)\approx 10^6\;{\rm pc}^{-3}$ and $\sigma(x_0)\sim 10^7\;{\rm cm}\;{\rm s}^{-1}$ \citep{Alexander05}. These values yield an incoming flux of stars at $r_1=2\times 10^{16}\;$cm of about $\dot{N}_s=4{\rm yr}^{-1}$, and a contribution of a few $10^{-7}\;{\rm yr}^{-1}$ to the collision rate with a cluster of $N_C=5000$ tightly captured stars. This value is two orders of magnitude smaller than the supply rate of tightly captured stars to the galactic center, and so will make a negligible impact on the hypervelocity collision rate. 

\section{The light curve following a hypervelocity collision}

A head-on, hypervelocity collision generates an outgoing shock wave that sweeps through the two stars, depositing both internal and kinetic energy. Strictly speaking, the source of energy in this scenario is external (the SMBH's gravity) rather than internal (e.g., thermonuclear burning or gravitational collapse of the core), but the deposited energy disrupts the stars ``explosively'', as in typical supernovae. Following the disruption, the gas expands while some thermal radiation escapes and produces a potentially observable signal. A full calculation of the light curve following a hypervelocity collision of two main sequence stars in the gravitational field of the SMBH 
goes beyond the scope of this work. Here we confine ourselves to a simple estimate of the time scales and characteristic luminosity. We crudely approximate the product of the collision as a spherical object with a radius $R_0$, which should be similar to the original radius of a single star. We farther assume that the shocked material has a uniform density and temperature, and a homologous velocity profile. These assumptions allow us to estimate the light curve with methods similar to those used for regular supernova (with one fine distinction, see below).  

As the explosion commences, the material is extremely opaque to its own thermal photons, and very little internal energy can escape by radiation from the surface. The hot matter expands adiabatically, cooling while converting internal energy into kinetic energy. Once the bulk of the internal energy has been converted, the material settles to a free streaming expansion. A total energy $E_0$ deposited in the material of two identical stars of mass $m_\star$ leads to an expansion speed of 
\begin{multline}\label{eq:v_exp}
v_{exp}=\frac{5}{3}\left(\frac{E_0}{m_\star}\right)^{1/2}=\\
1.2\times 10^9 \left(\frac{E_0}{10^{51}{\rm ergs}}\right)^{1/2}\left(\frac{m_\star}{M_\odot}\right)^{-1/2}{\rm cm}\;{\rm s}^{-1}\;.
\end{multline}
This is, naturally, of order the initial orbital velocity of the stars; the numerical factor in equation (\ref{eq:v_exp}) is appropriate for homologous expansion of a uniform density material. 

As long as the expansion is spherical, we can apply standard analyses of supernovae light curves, and, in particular, the ``radiative zero'' approximation, by \cite{ArnettBook}. We ignore the details of the early evolution where only a small fraction of the energy is released \citep[e.g.][]{NakarSari10}. Since the material is very opaque,  radiation leaks by diffusion,
and most of the energy leaks when the diffusion time is equal the expansion time with a typical luminosity , $L_0$, which is 
\begin{equation}\label{eq:L_0_RZ}
L_0 \cong {4 \pi} \frac{c}{\kappa}\left(\frac{R_\star}{2m_\star}\right)E_{th}\;.
\end{equation}
where $\kappa$ is specific opacity of the material, $E_{th}$ is the initial thermal energy in the explosion. 
An outgoing radial shock typically deposits about 1/2 of its energy as internal energy, so a 
reasonable estimate is that $E_{th}=0.5E_0$. With these assumptions, the typical luminosity $L_0$ for fully ionized material with solar composition comes out to be  
\begin{equation}\label{eq:L_0}
L_0=7.1\times 10^{39} 
\left(\frac{E_0}{10^{51} {\rm ergs}}\right)\left(\frac{\kappa}{0.34 {\rm cm}^2 {\rm g}^{-1}}\right)\;{\rm ergs}\;{\rm s}^{-1}\;.
\end{equation}
We note that equation (\ref{eq:L_0}) should include a factor of $R_\star/m_{\star}$ (compare to equation (\ref{eq:L_0_RZ})), but since this ratio is approximately constant for main sequence stars we omit it in the following, and use solar values as representative ones.    

In the particular case of a high energy collision between two sun-like stars the initial expansion time $t_H=R_\star/v_{exp}$ and the initial diffusion time $t_{diff}$,  are very different: the dynamical time is about sixty seconds, while the diffusion time scale is of order $10^{10}$ seconds. As the expansion time increases, and the diffusion time decreases, both linearly with the expansion radius or time, the two timescale becomes equal at $t_G=(2 t_H t_{diff,0})^{1/2}$,  
\begin{equation}\label{eq:t_Gauss}
t_G=2.4\times 10^6 \left(\frac{R_\star}{R_\odot}\right)^{1/2}
\left(\frac{m_\star}{M_\odot}\right)^{1/4}\left(\frac{E_0}{10^{51}{\rm ergs}}\right)^{-1/4}{\rm s}\;.
\end{equation}
after which the luminosity will decline with a Gaussian dependence on time.
Correspondingly, the light curve could roughly be described by a Gaussian with a typical time scale, $t_G$ of several days.  
However, well before the time scale of equation (\ref{eq:t_Gauss}), the evolution of the light curve is altered earlier by two competing effects.

{\it The recombination time.} 
During adiabatic expansion the temperature drops inversely with time. The expanding material ceases to be opaque when it cools to the hydrogen recombination temperature, $T_{rec}\approx10^4\;$K,
on a time scale of $t_{rec}=t_H\times(T_i/T_{rec})$, where $T_i$ is the initial temperature of the material after the explosion.
Again assuming $E_{th}=0.5E_0$, this initial thermal energy is dominated by radiation, and two identical stars which collide with a total energy of $E_0$ will lead to a recombination time of
\begin{equation}\label{eq:t_rec}
t_{rec}\approx 5.5 \left(\frac{E_0}{10^{51}{\rm ergs}}\right)^{-1/4}\left(\frac{m_\star}{M_\odot}\right)^{1/2}\left(\frac{R_\star}{R_\odot}\right)^{1/4}{\rm days}\;.
\end{equation}
Clearly, recombination will set in before the luminosity has been significantly affected by diffusion, and at recombination nearly all the remaining thermal energy will be emitted as the material becomes transparent.  
For an energy of $E_0=10^{51}\;$ergs and two sun-like stars, the remaining internal energy at recombination is about  $E_{th}(t_H/t_{rec}) \sim 6\times 10^{46}\;$ergs, comparable to the specific energy gained in hydrogen recombination, $Q_{rec}(H)=2.61\times 10^{46}$ergs per solar mass. If both internal and recombination energies are emitted over a time scale of order $t_{rec}$, the average luminosity during this recombination phase will be about $2\times 10^{41}\;{\rm ergs}\;{\rm s}^{-1}$, with a maximum a few times larger.

{\it The asphericity time.} 
Unlike regular supernovae, in our case the spherical approximation itself breaks down at a time of about $t_{asph}=r/v_{exp}$, where $r$ is the distance of the collision from the SMBH. At later times the SMBH's gravity causes the flow to become highly distorted, even if the explosion was initially spherical. Since $r\approx G M_\bullet m_\star/E_0$, the asphericity time scale is
\begin{equation}\label{eq:t_asph}
t_{asph}=10.5 \left(\frac{E_0}{10^{51}{\rm ergs}}\right)^{-3/2}\left(\frac{m_\star}{M_\odot}\right)^{3/2}\left(\frac{M_\bullet}{4\times 10^6 M_\odot}\right){\rm days}\;.
\end{equation}
In fact, this is nothing but the original orbital time of the binary (this follows from the definition of the tidal radius). Indeed for $a=0.1AU$, the orbital time is 10 days.

For two sun-like stars and a total energy of $E_0=10^{51}\;$ergs the recombination time scale is shorter than the aspherecity time, and the estimate given above for the luminosity at the recombination peak should hold. However, the situation is reversed for $E_0$ of just a few $10^{51}\;$ergs. For example, in a collision between two sun-like stars at the binary tidal disruption radius mentioned above, $r_T=2.4\times 10^{14}\;$cm, the total energy is about $4.4\times 10^{51}\;$ergs. In this case the flow becomes aspherical at $t_{asph}\approx 1.1\;$days, while the recombination time scale only reduces to $t_{rec}\approx 3.8\;$days. Hence, energetic collisional supernovae become aspherical prior to the onset of recombination. Once the SMBH's gravity dominates the flow, the remaining internal energy does not change significantly until the material becomes transparent. For $E_0=4.4\times 10^{51}\;$ergs, this energy is about $6.4\times 10^{47}\;$ergs, and if emitted over a time scale of order $t_{asph}$, the resulting luminosity will be several $10^{42}\;{\rm ergs}\;{\rm s}^{-1}$.

\section{A comment about late time accretion-induced emission}

A collisional supernova should serve as a prelude to a later stage of observable emission, once the ejected material begins to interact with the SMBH. The general features in this stage should be similar to the outcome in a single star tidal disruption event, TDE \citep{Rees88}. Specifically, most (or even all) of the stellar matter remains bound to the SMBH, and must fall back towards it, eventually forming an accretion disk, leading to X-ray emission \citep{Rees88,EvansKochanek89}. Furthermore, as the debris engulfs the SMBH it shocks onto itself, reheating the material and generating a flare, mostly in ultraviolet and optical wavelengths \citep{StrubbeQuateart09}. Different elements of the debris shock at different times, but the general time scale should be several $t_{asph}$. This implies a supply rate of matter for accretion which is initially larger than the Eddington limit, $\dot{M}_{Edd}\approx 2\times 10^{-2} M_\odot(M_{\bullet}/10^6 M_\odot)\;{\rm yr}^{-1}$ (assuming a $10\%$ efficiency of converting accreted matter into radiation). Super-Eddington accretion will lead to outgoing mass flow, and an extended period of ultraviolet and optical emission with a power of about $L_{Edd}\approx 1.3\times 10^{44}(M_\bullet/10^6\;M_\odot)\;$ergs s$^{-1}$, alongside with soft X-ray emission from the accretion disk. 

We do not attempt to quantify the emission at this stage, since the details are likely to be dependent on the initial conditions. In principle, both the total mass of the gas that remains bound to the SMBH and the distribution of the arrival time scales of this gas can vary significantly as a function of the combination of explosion energy and the velocity of the center of mass. For example, a head on collision with a zero center of mass velocity will eject a fraction of the mass in a ballistic trajectory towards the SMBH, while in the other extreme of a collision with a very high center of mass velocity, practically all of the ejecta will initially orbit the SMBH, very similar to the dynamics of tidal disruption of a single star. Numerical simulations are required for a comprehensive analysis (such as recently applied by \citet*{HSL12} in the context of TDE's). We generally conclude that a collisional supernova is destined to be followed by a longer, more powerful, second act as the debris interacts with the SMBH. Our analysis suggests that higher energy disruptions generate light curves that transform continuously from the explosion to this extended emission, which becomes dominant after several $t_{asph}$. Lower energy explosions with $t_{rec}\leq t_{asph}$ may have a short dim interval between the supernova and the extended emission. No presently known TDE candidates have been observed early enough to allow for an identification of an initial supernova phase, but we suggest that some of these candidates may actually be the result of a hypervelocity stellar collision. 

\section{Observational signature}

The optical emission from a collisional-supernova is equivalent to
dimmer core collapse supernovae; for example, the maximum luminosity
of SN1987A was about $10^{42}\;$ergs s$^{-1}$. The time scale is only
several days, which is an order of magnitude shorter than standard
supernovae. In terms of power and time scales, collisional supernovae
are similar to the so-called ``Type .Ia'' supernovae, such as SN2002bj
and SN2010X, which are believed to result from violent helium ignition
on the surface of an accreting white dwarf
\citep{pointIa1,pointIa2}. Nonetheless collisional supernovae should
be readily distinguishable from all other types of supernovae, both
standard and dim and fast explosions. First, they will be found only
at the centers of their host galaxies, and will be followed, as
mentioned above, by a longer stage of accretion induced
emission. Second, in the absence of thermonuclear reactions of heavier
elements, their light curves have no (exponential) tail from
radioactive decays. Finally, the ejected material is mostly hydrogen
moving at a speed of $\sim 10^{4}\;{\rm km}\;{\rm s}^{-1}$, whereas
standard hydrogen dominated (Type II) supernovae typically eject their
envelopes at lower speeds. The high speed of the hydrogen gas could be
inferred from an observed spectrum of a sufficiently nearby (and hence
bright) transient.

In order to identify a fast supernova near a galactic center, dedicated searches are required. Collisional supernovae will have peak magnitudes $M\approx -17\;{\rm to} -15$ and should be observable in other galaxies if not confused by other bright sources or obscured by significant dust extinction. However, they must also be distinguished from TDEs, which are also expected at a general rate of $10^{-5}-10^{-4}\;$yr$^{-1}$ per galaxy \citep{TDErate}. Such an observational distinction will depend on capturing the light curve prior to the late-time emission, so relatively rapid cadence is required: about once per day. This necessitates transient surveys such as the Panoramic Survey Telescope and Rapid Response System (Pan-STARRS), the Palomar Transient Factory (PTF) and the Large Synoptic Survey Telescope (LSST). Conservatively assuming that positive detection of events near galactic centers requires an apparent magnitude of $m_B\geq 20$, a detection range of $D=200$Mpc should be reasonable. If the rate of collisional supernovae per galaxy is $10^{-5}\;{\rm yr}^{-1}$, PTF observations with daily cadences to a depth of $m_B\geq 20$ could detect the supernova phase in a few events annually (with a ten-fold increase for future surveys such as Pan-STARRS-4, and LSST). It would be prudent to conduct a careful analysis of every TDE candidate which is caught early enough in its evolution, and examine it for the signature of a collisional supernova.

\section*{Acknowledgments}
We thank Orly Gnat and Eran Ofek for useful discussions and comments. This work supported in part by ERC and ISF grants and a Packard fellowship. (R.S.), and by NSF grant AST-0907890 and NASA grants NNX08AL43G and NNA09DB30A (A.L.).

\end{document}